\documentclass[10pt,conference]{IEEEtran}
\usepackage[latin1]{inputenc}
\usepackage[numbers, square, comma, compress]{natbib}
\usepackage[english]{babel}
\usepackage{dsfont}
\usepackage{setspace}
\usepackage{amsfonts}
\usepackage{amssymb}
\usepackage{amsmath}
\usepackage{mathrsfs}
\usepackage{xcolor}
\usepackage{graphicx}
\usepackage{mathdots}
\usepackage{float}
\usepackage{dblfloatfix}
\usepackage{stmaryrd}
\usepackage{epstopdf}
\usepackage[colorinlistoftodos]{todonotes}
\usepackage[ruled,vlined]{algorithm2e}
\usepackage{verbatim}

\usepackage{tikz}
\usetikzlibrary{shapes,arrows}
\usetikzlibrary{calc}

\pdfoutput=1

\newtheorem{teo}{Theorem}

\newtheorem{lem}{Lemma}

\newtheorem{oss}{Remark}
\newtheorem{defi}{Definition}

\DeclareMathOperator{\X}{\mathcal X}

\DeclareMathOperator{\h}{\mathcal H}
\DeclareMathOperator{\V}{\mathcal V}

\IEEEoverridecommandlockouts

\title{Polar Coding for Empirical Coordination of Signals and Actions over Noisy Channels}

\author{\IEEEauthorblockN{Giulia Cervia\IEEEauthorrefmark{1}, Laura Luzzi\IEEEauthorrefmark{1}, Matthieu R. Bloch\IEEEauthorrefmark{3},  and Ma\"{e}l Le Treust\IEEEauthorrefmark{1}  \thanks{The work of M.R. Bloch was supported in part by NSF under grant CIF 1320304. The work of M. Le Treust was supported by INS2I CNRS through project PEPS JCJC CoReDe 2015 and PEPS INS2I StrategicCoo 2016.}}
\IEEEauthorblockA{\IEEEauthorrefmark{1}Laboratoire ETIS (ENSEA - UCP - CNRS), Cergy-Pontoise, France \\
Email: \{giulia.cervia, laura.luzzi, mael.le-treust\}@ensea.fr}
\IEEEauthorblockA{\IEEEauthorrefmark{3}School of Electrical and Computer Engineering, Georgia Institute of Technology, Atlanta, Georgia\\
Email: matthieu.bloch@ece.gatech.edu} 
}

\begin{document}
\bstctlcite{IEEEexample:BSTcontrol}
\IEEEoverridecommandlockouts
\maketitle

 \begin{abstract} We develop a polar coding scheme for empirical coordination in a two-node network with a noisy link in which the input and output signals have to be coordinated with the source and the reconstruction. In the case of non-causal encoding and decoding, we show that polar codes achieve the best known inner bound for the empirical coordination region, provided that a vanishing rate of common randomness is available. This scheme provides a constructive alternative to random binning and coding proofs. \end{abstract}

\section{Introduction}
Coordinating behavior in decentralized networks is a fundamental challenge for many applications, such as cognitive radio, autonomous vehicles, cloud computing and smart grids. These networks are composed of autonomous devices that sense their environment and choose their actions in order to achieve a general objective. Within the framework of information theory, the problem of coordination has been investigated in \cite{cuff2010} and two different metrics have been proposed to measure the level of coordination. Empirical coordination requires the joint histogram of the actions to approach a target distribution, while strong coordination requires the total variation distance of the distribution of actions to converge to an i.i.d. target distribution. Explicit schemes using polar codes for point-to-point coordination have been proposed in the case of empirical coordination uniform actions  \cite{blasco-serrano2012}, strong coordination for uniform actions \cite{bloch2012strong} and then generalized to the case of non uniform actions \cite{chou2015coordination}. In all these works the communication links are assumed to be error-free.

In this paper we consider a two-node network with an information source and a noisy channel. We focus on the setting in which both the encoder and the decoder are non-causal. Coordination in state-dependent networks with different observation hypotheses (causal and strictly causal encoder/decoder) has been studied in \cite{larrousse2015coordinating,treust2014correlation,treust2015empirical}. Following the framework in \cite{treust2015empirical,cuff2011hybrid,treust2014empirical}, we require empirical coordination of the channel input and output signals with the source and the reconstruction. This requirement allows us to consider scenarios in which the actions performed by an agent play a double role, influencing the global behavior, as well as carrying information for the other agents \cite{gossner2006optimal,cuff2011coordination,larrousse2013coded}. In \cite{cuff2011hybrid} the authors provide an inner bound for the set of achievable joint empirical distributions, called the \emph{coordination region}. This is done by considering the situation as a joint source-channel problem in which the channel inputs are coordinated with the source symbols and decoder outputs. This scenario, in which signals and actions are coordinated, can be applied to watermarking, coded power control \cite{larrousse2015coordination} and general decentralized networks in which devices observe signals and choose actions. 

Inspired by the binning technique using polar codes in \cite{chou2015polar}, we propose an explicit polar coding scheme that achieves the inner bound for the coordination capacity region in \cite{cuff2011hybrid} by using a negligible amount of common randomness. We use a chaining construction as in \cite{hassani2014universal,mondelli2015achieving} to ensure proper alignment of the polarized sets.

The remainder of the paper is organized as follows. $\mbox{Section \ref{sec:problem}}$ introduces the notation, describes the model under investigation and states the main achievability result. $\mbox{Section \ref{sec:polar}}$ details the proposed coordination scheme using polar codes. Finally, Section \ref{subsec:coord} proves the main result.

\begin{center}
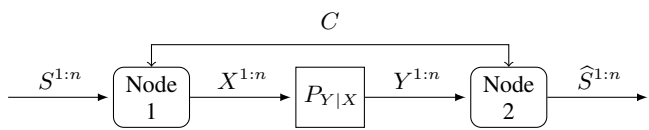
\begin{figure}[b]
\begin{small}
\begin{tikzpicture}[
nodetype1/.style={
	rectangle,
	minimum width=7mm,
	text width=7mm,
	text centered,
	minimum height=8mm,
	draw=black
},
nodetype2/.style={
	rectangle,
	rounded corners,
	minimum width=8mm,
	text width=8mm,
	text centered,
	minimum height=8mm,
	draw=black
},
tip2/.style={-latex,shorten >=0.6mm}
]
\matrix[row sep=0.4cm, column sep=1.4cm, ampersand replacement=\&]{
\node ( ) {}; \& 
\node ( ) {}; \& 
\node (C) {$C$}; \& \\
\node (PS)  {}; \& 
\node (uno) [nodetype2, text centered] {Node \\ 1}; \& 
\node (PYX) [nodetype1] {$P_{Y|X}$}; \& 
\node (due) [nodetype2, text centered] {Node \\ 2}; \&
\node (PST) {}; \\};
\draw[<->] (uno)  -- ++ (0,0.7) -|  (due) ;
\draw[->] (PS) edge[tip2] node [above] {$S^{1:n}$} (uno) ;
\draw[->] (uno) edge[tip2] node [above] {$X^{1:n}$} (PYX) ;
\draw[->] (PYX) edge[tip2] node [above] {$Y^{1:n}$} (due) ;
\draw[->] (due) edge[tip2] node [above] {$\widehat S^{1:n}$} (PST) ;

\end{tikzpicture}
\end{small}
\vspace{-0.4cm}
\caption{Coordination of signals and actions for a two-node network with a noisy channel.}
\label{fig: coord}
\end{figure}
\end{center}

\begin{center}
\begin{figure*}[!t]
\begin{small}
\begin{tikzpicture}[
nodetype1/.style={
	rectangle,
	minimum width=10mm,
	text width=10mm,
	text centered,
	minimum height=8mm,
	draw=black
},
nodetype2/.style={
	rectangle,
	rounded corners,
	minimum width=11mm,
	text width=11mm,
	text centered,
	minimum height=8mm,
	draw=black
},
nodetype3/.style={
	rectangle,
	minimum width=10mm,
	text width=10mm,
	text centered,
	minimum height=8mm,
	draw=black,
    dashed
},
tip2/.style={-latex,shorten >=0.6mm}
]
\matrix[row sep=0.3cm, column sep=1.6cm, ampersand replacement=\&]{
\node( ) {}; \& 
\node( ) {}; \& 
\node(C) {$C$}; \& \\
\node (Source)   {}; \& 
\node (SourceEncoder) [nodetype2, text centered] {Enc.}; \& 
\node (f) [nodetype3] {$P_{X|US}$}; \& 
\node (Channel) [nodetype1] {$P_{Y|X}$}; \&
\node (ChannelDecoder) [nodetype2] {Dec.}; \&
\node (g) [nodetype3]  {$P_{\widehat S|UY}$}; \&
\node (Reconstruction) {};\\};
\draw[<->] (SourceEncoder)  -- ++ (0,0.7) -|  (ChannelDecoder) ;
\draw[->] (Source) edge[tip2] node [above] {$S^{1:n}$} (SourceEncoder) ;
\draw[->] (SourceEncoder) edge[tip2] node [above] {$\widetilde{U}^{1:n}$} (f) ;
\draw[->] (f) edge[tip2] node [above] {$X^{1:n}$} (Channel) ;
\draw[->] (Channel) edge[tip2] node [above] {$Y^{1:n}$} (ChannelDecoder) ;
\draw[->] (ChannelDecoder) edge[tip2] node [above] {$\widehat{U}^{1:n}$} (g) ;
\draw[->] (g) edge[tip2] node [above] {$\widehat{S}^{1:n}$} (Reconstruction) ;

\end{tikzpicture}
\end{small}
\vspace{-0.1cm}
\caption{Joint source-channel model. Although we require common randomness $C$, we show that the rate is negligible.}
\label{fig: schema}
\end{figure*}
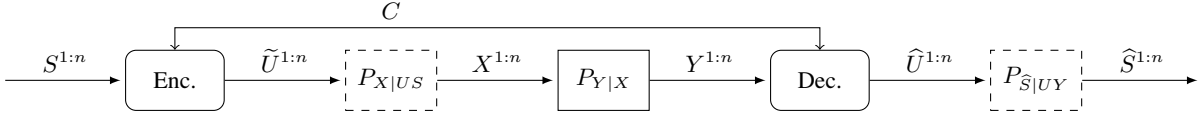
\end{center}

\vspace{-1.5cm}
\section{Problem statement}\label{sec:problem}
\subsection{Notation}\label{sec: not}
We define the integer interval $[a,b]$ as the set of the integers between $a$ and $b$.
For $n=2^m$, $m \in \mathbb N$, we note $G_n:= \begin{footnotesize}
\begin{bmatrix}
1 & 0\\
1 & 1
\end{bmatrix}^{\otimes m}
\end{footnotesize}$ the source polarization transform defined in \cite{arikan2010source}.
Given $X^{1:n}:= (X^1, \ldots, X^n)$ a random vector, we note $X^{1:j}$ the first $j$ components of $X^{1:n}$ and $X[A]$, where $A \subset [1,n]$, the components $X^j$ such that $j \in A$.
We note $\mathbb V (\cdot , \cdot)$ and $\mathbb D (\cdot \Arrowvert \cdot)$ the variational distance and the Kullback-Leibler divergence between two distributions, respectively. We note $T_{X^{1:n}}$ the empirical distribution of a random vector $X^{1:n}$ taking values in $\X^n$. Given a distribution $P_X$, $X^{1:n}$ is in the $\epsilon$-typical set $T_{\epsilon} (X)$ if
$\mathbb V (T_{X^{1:n}},P_X) \leq \epsilon.$

\subsection{System model and main result}\label{sec: model}
We start with the model depicted in Figure \ref{fig: coord} and consider two agents, Node 1 and Node 2, who have access to a shared randomness source $C \in \mathcal C_n$. Node 1 draws an i.i.d. sequence of actions  $S^{1:n} \in \mathcal S^n$ according to a discrete probability distribution $P_S$. Node 1 then selects a signal $X^{1:n}= f_n(S^{1:n}, C)$, where $f_n: \mathcal S^n \times  \mathcal C_n \rightarrow \X^n$ is the non-causal encoder. The signal $X^{1:n}$ is transmitted over a discrete memoryless channel parametrized by the conditional distribution $P_{Y |X} $.
Upon receiving $Y^{1:n} \in \mathcal Y^n $, Node 2 selects an action $\widehat S^{1:n} = g_n(Y^{1:n}, C)$, where $g_n: \mathcal Y^n  \times  \mathcal C_n \rightarrow \widehat{\mathcal S}^n$ is the non-causal decoder.
For block length $n$, the pair $(f_n , g_n )$  constitutes a code.
Node 1 and Node 2 wish to coordinate in order to obtain a joint distribution of actions and signals that is close to a target distribution $P_{SXY\widehat{S}}$.
We focus on the empirical coordination metric defined in \cite{cuff2010}.
\begin{defi}
A distribution $P_{SXY\widehat{S}}$ is \emph{achievable} if for all 
$\epsilon >0$ there exists a code $(f_n, g_n)$ 
such that
$$\lim_{n\rightarrow \infty}\mathbb P \left\{ \mathbb V \left(  T_{S^{1:n} X^{1:n} Y^{1:n} \widehat{S}^{1:n}},P_{SXY\widehat{S}} \right) > \epsilon \right\} =0,$$
where $T_{S^{1:n} X^{1:n} Y^{1:n} \widehat{S}^{1:n}}(s,x,y,\hat s)$ 
is the empirical distribution of the tuple $(S^{1:n}, X^{1:n}, Y^{1:n}, \widehat{S}^{1:n})$ 
induced by the code. The \emph{empirical coordination region} $\mathcal R$ is the set of achievable distributions $P_{SXY\widehat{S}}$.
\end{defi}

In the case of non-causal encoder and decoder, the problem of characterizing the empirical coordination region is still open, but the following inner bound was proved in \cite{cuff2011hybrid}.
\begin{teo} \label{teocuff}
Let $P_{S}$ and $P_{Y|X}$ be the given source and channel parameters. When the encoder and decoder are allowed to be non-causal, the region $\mathcal R'\subset \mathcal R$ defined below is included in the empirical coordination region.
\begin{equation}\label{eq: c}
\mathcal R' := \begin{Bmatrix} P_{SXY\widehat{S}} \mbox{ } : \mbox{ }\exists U \mbox{ taking values in $\mathcal U$ s.t. }\\ P_{SXY\widehat{S}U}= 
P_S P_{U|S} P_{X|US} P_{Y|X} P_{\widehat{S}|UY},\\\mbox{ }I(U;S) \leq I(U;Y),\\ \mbox{ } \lvert \mathcal U \rvert \leq \lvert \mathcal S \rvert \lvert \mathcal X \rvert \lvert \mathcal Y \rvert \lvert \widehat {\mathcal S} \rvert +1 \end{Bmatrix}.
\end{equation}
\end{teo}

We propose a scheme based on polar coding that achieves the inner bound $\mathcal R'$ for the empirical coordination region. The key step for coordination is to generate the same auxiliary sequence $U^{1:n}$ at
the decoder and the encoder.  Once this is accomplished, the task is essentially done because the sequences $X^{1:n}$ and $Y^{1:n}$ with the correct distribution can be generated via the conditional distributions $P_{X|US}$ and the channel $P_{Y|X}$; hence, the appropriate $\widehat{S}^{1:n}$ can be drawn at the decoder. For brevity, we only focus on the set of achievable distributions in $\mathcal R'$ for which the auxiliary variable $U$ is binary. The scheme can be generalized to the case of a non-binary random variable $U$ using non-binary polar codes. We now state the main result of the paper.
\begin{teo}\label{theo}
For all $P_{SXY\widehat{S}}$ for which there exists $U$ taking values in $\mathcal U = \{ 0,1\}$ such that 
$$P_{SXY\widehat{S}U}= 
P_S P_{U|S} P_{X|US} P_{Y|X} P_{\widehat{S}|UY},$$ 
there exists an explicit polar coding scheme that achieves empirical coordination with rate of common randomness $\frac{\log_2 \lvert \mathcal C_n \rvert}{n}$ that goes to zero as $n$ goes to infinity.
\end{teo}

\section{Polar coding for coordination of signals and actions}\label{sec:polar}

\subsection{Polar coding scheme}\label{subsec:scheme}
We suppose that $P_{SXY \widehat{S}}$ belongs to $\mathcal R'$ and show how to achieve empirical coordination with polar codes. 

Consider the random vectors $S^{1:n}$, $U^{1:n}$, $X^{1:n}$, $Y^{1:n}$ and $\widehat S^{1:n}$
generated i.i.d. according to $P_{SXUY \widehat S}$ that satisfies \eqref{eq: c}.  Let $V^{1:n}=U^{1:n}G_n$ the polarization of $U^{1:n}$,  where $G_n$ is the source polarization transform defined in Section \ref{sec: not}.
For some $0<\beta<1/2$, let $\delta_n = 2^{-n^ {\beta}}$ and define the very high entropy and high entropy sets:
 \begin{align}\label{eq: hv}
 \begin{split}
\V_{V}: & =\left\{j\in[1;n]:H(V^j|V^{1:j-1})>1-\delta_n \right\},\\
\V_{V | S}: & =\left\{j\in[1;n]:H(V^j|V^{1:j-1} S^{1:n})>1-\delta_n \right\}, \\
\V_{V | Y}: & =\left\{j\in[1;n] : H(V^j|V^{1:j-1} Y^{1:n})>1-\delta_n \right\},\\
\h_{V | Y}: & =\left\{j\in[1;n]:H(V^j|V^{1:j-1} Y^{1:n})>\delta_n \right\} .
\end{split}
 \end{align}

Now define the following sets:
\begin{align*}
A_1 := \mathcal V_{V|S} \cap \mathcal H_{V|Y} , \quad & A_2 : = \mathcal V_{V|S} \cap \mathcal H_{V|Y}^c,\\
A_3 := \mathcal V_{V|S}^c \cap \mathcal H_{V|Y} ,\quad
& A_4 := \mathcal V_{V|S}^c \cap \mathcal H_{V|Y}^c.
\end{align*}

\begin{oss}\label{oss card}
We have:
\begin{itemize}
\item $\V_{V | Y} \subset \h_{V | Y}$ and $ \lim_{n \rightarrow \infty} \frac{\lvert \h_{V | Y} \setminus \V_{V|Y} \rvert}{n} = 0$ \cite{arikan2010source},
 \item  $\lim_{n \rightarrow \infty} \frac{\lvert \V_{V|S} \rvert}{n}  = H(U|S)$ \cite{chou2015secretkey},
 \item  $ \lim_{n \rightarrow \infty} \frac{\lvert \h_{V | Y} \rvert}{n} = H(U|Y)$ \cite{arikan2010source}.
\end{itemize}
Since  $H(U|S)- H(U|Y) = I(U;Y)- I(U;S),$ for sufficiently large $n$ 
the assumption $I(U;Y) \geq I(U;S)$ implies directly that $\lvert A_2 \rvert \geq \lvert A_3 \rvert$.
\end{oss}
\subsection{Encoding}\label{subsec:enc}
Note that the set $A_3$ is non-empty in general. The bits $V^j$ with $j \in A_3$ can be generated at the encoder according to the previous bits, but cannot be recovered reliably at the decoder. 
To overcome this issue, we code over multiple blocks and use a chaining construction as in \cite{chou2015polar}. The encoder observes $k$ blocks of the source $(S_1^{1:n}, \ldots, S_k^{1:n})$ and 
generates for each block $i\in \{1, \ldots,k\}$ a random variable $\widetilde V_i^{1:n}$ following the procedure described in Algorithm \ref{alg1}.
\begin{center}\label{fig:kblo}
\begin{figure}[ht!]
 \centering
 \includegraphics[scale=0.6]{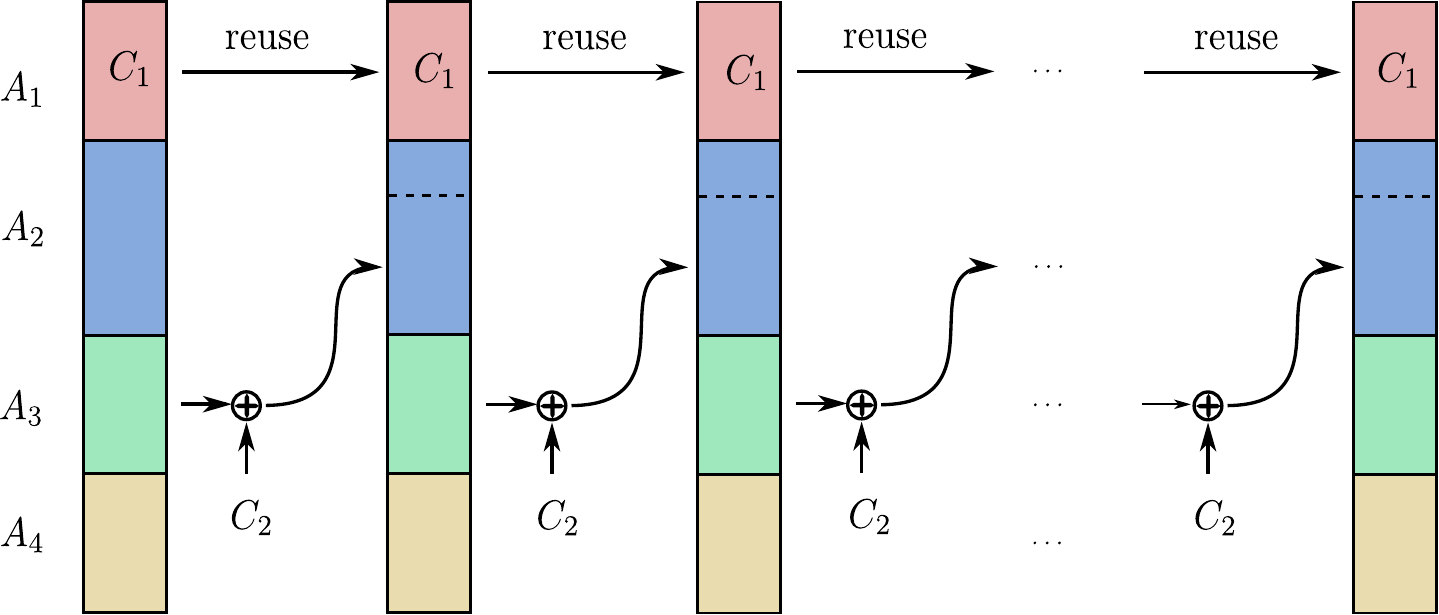}
\caption{Chaining construction for block Markov encoding}
\end{figure}
\end{center}
\vspace{-1cm}

\begin{algorithm}[ht!]\label{alg1}
\begin{small}
\DontPrintSemicolon
\SetAlgoLined
\SetKwInOut{Input}{Input}
\SetKwInOut{Output}{Output}
\Input{ $(S_1^{1:n}, \ldots, S_k^{1:n})$, $M$ local randomness (uniform random bits) and 
common randomness $C=(C_1, C_2)$ shared with Node 2:
$C_1$ of size $\lvert A_1 \rvert$ and $C_2$ of size $\lvert A_3 \rvert$.
}

\Output{$\left( \widetilde V^{1:n}_1, \ldots, \widetilde V^{1:n}_k \right)$}
\If{$i=1$}{
 $\widetilde V_{1}[A_1] \longleftarrow C_1 \qquad \widetilde V_{1}[A_2]  \longleftarrow M $\;
\For{$j \in A_{3} \cup A_{4}$}{
Given $S_1^{1:n}$, succ. draw the bits $\widetilde V_{1}^j$ according to 
\begin{equation} \label{eq: p1}
P_{V^j \mid V^{1:j-1} S^{1:n}} \left(\widetilde V_{1}^j\mid \widetilde V_1^{1:j-1}S_1^{1:n} \right)
\end{equation}
}
}

\For{$i=2, \ldots, k$}{
 $\widetilde V_{i}[A_1] \longleftarrow C_1 \qquad \widetilde V_{i}[A'_3] \longleftarrow \widetilde V_{i-1}[A_{3}] \oplus C_2$\;
 $\widetilde V_{i}[A_2 \setminus A'_3] \longleftarrow M $ \;
\For{$j \in A_{3} \cup A_{4}$}{
Given $S_i^{1:n}$, succ.  draw the bits $\widetilde V_{i}^j$ according to
\begin{equation} \label{eq: p2}
P_{V^j \mid V^{1:j-1}S^{1:n}} \left(\widetilde V_{i}^j \mid \widetilde V_i^{1:j-1}S_i^{1:n} \right)
\end{equation} 
}
}
\BlankLine
\caption{Encoding algorithm at Node 1}
\end{small}
\end{algorithm}

In particular, the chaining construction proceeds as follows:
\begin{itemize}
\item since the bits in $\V_{V|S}$ are nearly uniform and independent of $S^{1:n}$ by Definition \eqref{eq: hv}, the bits in  $A_1 \subset \V_{V|S}$ are chosen with uniform probability using a uniform randomness source $C_1$ shared with Node 2, and their value is reused over all blocks;
\item in the first block the bits in  $A_2 \subset \V_{V|S}$ are chosen with uniform probability using a local randomness source $M$;
\item for the  following blocks, let $A'_3$ be a subset of $A_2$ such that $\lvert A'_3 \rvert= \lvert A_3 \rvert$.  The bits of $A_3$ in block $i$ are sent to $A'_3$ in the block $i+1$ using a one time pad with key $C_2$. Thanks to the Crypto Lemma 
 \cite[Lemma 3.1]{bloch2011physical}, if we choose $C_2$ of size $\lvert A_3 \rvert$ to be a uniform random key, the bits in $A'_3$ in the block $i+1$ are uniform. The bits in $A_2 \setminus A'_3$ are chosen with uniform probability using the local randomness source $M$;
\item the bits in $A_3$ and in $A_4$ are generated according to the previous bits using successive cancellation encoding \cite{arikan2010source}. Note that it is possible to sample efficiently from $P_{V^i \mid V^{1:i-1} S^{1:n}}$ given $S^{1:n}$ \cite{arikan2010source}.
\end{itemize}

The encoder then computes $\widetilde U_i^{1:n}=\widetilde V_i^{1:n} G_n$ for $i=1, \ldots, k$ and generates $X_i^{1:n}$ symbol by symbol from $\widetilde U_i^{1:n}$ and $S_i^{1:n}$ using the conditional distribution $$P_{X_i^j |\widetilde U_i^j  S_i^j}(x|\widetilde u_i^j, s_i^j)=P_{X|US} ( x | u_i^j, s_i^j)$$ and sends $X_i^{1:n}$ over the channel.

We use an extra $(k+1)$-th block to send 
a version of $V_k[A_{3}]$ encoded with a good channel code. In particular, this can be done using the polar code construction for asymmetric channels stated in \cite{honda2013polar}. Let $Z^{1:n}= X^{1:n} G_n $ be the polarized version of $X^{1:n}$. We place the information $V_k[A_{3}]$ in the positions of $Z^{1:n}$ indexed by $\V_X \cap \h^c_{X \mid Y}$. We note that $\V_X \cap \h^c_{X \mid Y}$ has cardinality approximately equal to $n  I(X;Y)$ \cite{honda2013polar}.
We have $\lvert A_3 \rvert \leq \lvert A_2 \rvert \leq \lvert \V_V \cap \h^c_{V \mid Y}\rvert$,  which is approximately $n  I(U;Y)$.
By hypothesis, we have the Markov chain $U-X-Y$ and therefore  $\lvert A_3 \rvert \leq n I(X;Y) $. We can send the bits in $A_3$ with vanishing error probability. 
The scheme in \cite{honda2013polar} requires common randomness, which will
have vanishing rate when $k$ is large enough since it's used only in the last block, and uniform messages, which can be achieved using a one-time-pad as before. Finally, $\widetilde{X}_{k+1}^{1:n}$ is the output of the channel code described above.

\subsection{Decoding}\label{subsec:dec}
\begin{algorithm}[ht!]\label{alg2}
\begin{small}
\DontPrintSemicolon
\SetAlgoLined

\SetKwInOut{Input}{Input}
\SetKwInOut{Output}{Output}
\Input{$(Y_1^{1:n}, \ldots, Y_{k+1}^{1:n})$, $C=(C_1, C_2)$ common randomness shared with Node 1}
\Output{$(\widehat V_1^{1:n}, \ldots, \widehat V_{k}^{1:n})$}
\For{$i=k, \ldots, 1$}{
 $\widehat V_i[A_1] \longleftarrow C_1$\;
 \If{$i=k$}{$\widehat V_{i}[A_3] \longleftarrow Y_{k+1}^{1:n}$ as in \cite{honda2013polar}}
 \Else{$\widehat V_{i}[A_3] \longleftarrow \widehat V_{i+1} [A'_{3}]$\;}
\For{$j \in A_{2} \cup A_{4}$}{ Successively draw the bits according to
$$\widehat V_i^j = \begin{cases}
 0 \quad \mbox{if } L_n(Y_i^{1:n}, V_i^{1:j-1}) \geq 1\\
 1 \quad \mbox{else} 
 \end{cases}$$
\begin{footnotesize}
$$L_n(Y_i^{1:n}, V_i^{1:j-1}) = \frac{P_{V_i^j \mid V_i^{1:j-1}  Y_i^{1:n}}\left(0 \mid \widehat V_i^{1:j-1} Y_i^{1:n} \right) }{P_{V_i^j \mid V_i^{1:j-1}  Y_i^{1:n}}\left(1 \mid \widehat V_i^{1:j-1}Y_i^{1:n} \right)}$$
\end{footnotesize}}
}
\BlankLine
\caption{Decoding algorithm at Node 2}
\end{small}
\end{algorithm}
The decoder observes $(Y_1^{1:n}, \ldots, Y_{k+1}^{1:n})$ and 
the $(k+1)$-th block allows it to decode in reverse order.
In block $i \in [1,k]$, the decoder has access to $\widehat V_i[A_{1} \cup A_{3}]= \widehat  V_i[\h_{V \mid Y}]$:
\begin{itemize}
\item the bits in $A_1$ correspond to shared randomness $C_1$;
\item in block $k$, the bits in $A_3$ are recovered from $Y_{k+1}^{1:n}$ using the decoding process in \cite{honda2013polar};
\item in block $i \in [1, k-1]$ the bits in $A_3$ are obtained by successfully recovering $A_2$ in block $i+1$.
\end{itemize}
For each block $i=k, \ldots, 1$ the decoder recovers an estimate $\widehat V_i^{1:n}$ of $\widetilde V_i^{1:n}$ using Algorithm \ref{alg2}. From ${Y_i}^{1:n}$ and $\widehat V_{i}[A_1 \cup A_{3}]$ the successive cancellation decoder can retrieve $\widehat V_i[A_2 \cup A_4]$.
Note that as shown in \cite[Theorem 3]{arikan2010source}, we have:
\begin{equation}\label{eq: ar}
\lim_{n \rightarrow \infty}\mathbb P \left\{ \widetilde{V}^{1:n}= \widehat V^{1:n}\right\}=1.
\end{equation}
The decoder computes $\widehat U_i^{1:n} = \widehat V_i^{1:n} G_n $. Then it generates $\widehat S_i^{1:n}$ symbol by symbol using: $$P_{\widehat S_i^j | \widehat U_i^j Y_i^j} (s|u,y)=P_{\widehat S | UY}(s|u,y).$$

\begin{oss}
The encoding and decoding complexity of this scheme is $O \left( n k \log n \right)$.
\end{oss}

\subsection{Rate of common randomness}\label{subsec:rate}
The rate of common randomness $C$ is  negligible since: $$\lim_{n \rightarrow \infty \atop k \rightarrow \infty} \frac{\lvert A_1 \cup A_3 \rvert}{kn}=\lim_{n \rightarrow \infty \atop k \rightarrow \infty} \frac{\lvert \h_{V|Y}\rvert}{kn} = \lim_{k \rightarrow \infty} \frac{H(U|Y)}{k} =0.$$

\section{Proof of Theorem \ref{theo}}\label{subsec:coord}
\subsection{Preliminary results}\label{subsec: lemmas}
We first state a few lemmas that we will need to prove Theorem \ref{theo}. The proofs can be found in the Appendix.

\begin{lem}\label{coo1}
 For any $i \in [1,k]$, for all $\epsilon_0 >0$, $$\lim_{n \rightarrow \infty} \mathbb P \left\{ \mathbb{V}\left( T_{S^{1:n}_{i} \widetilde U^{1:n}_{i}}, P_{SU}\right) > \epsilon_0 \right\} =0.$$
\end{lem}

\begin{lem}\label{lem3}
Let $P_A$ a distribution, $A^{1:n}$ a random vector,
$B^{1:n}$ a random vector generated from $A^{1:n}$ with i.i.d. conditional distribution $P_{B|A}$ and suppose
$\displaystyle \lim_{n\rightarrow \infty} \mathbb P\left\{\mathbb V\left(T_{A^{1:n}}, P_A\right)> \epsilon \right\} =0.$
Then, for all $\epsilon'>\epsilon$ we have:
$$ \lim_{n\rightarrow \infty} \mathbb P\left\{\mathbb V\left( T_{A^{1:n} B^{1:n}}, P_{AB}\right) > \epsilon'\right\} =0.$$
\end{lem}

\begin{lem}\label{vsp}
 Let $X^{1:n}$, $\widetilde X^{1:n}$ two possibly dependent random sequences taking values in $\X^{n}$ and define
 $$T_{\left(X^{1:n},\widetilde X^{1:n}\right)}(x):= \frac{1}{2n} \sum_{i=1}^n \left( \mathds 1 \{X^i=x\} + \mathds 1 \{\widetilde X^i=x\}\right).$$ 
Then for any distribution $P$  on $\X$,
\begin{equation*}
 \mathbb{V}\left(T_{\left(X^{1:n},\widetilde X^{1:n}\right)}, P\right) \leq \frac{1}{2} \mathbb{V}\left(T_{X^{1:n}}, P\right) + \frac{1}{2} \mathbb{V}\left(T_{\widetilde X^{1:n}}, P\right).
\end{equation*}
\end{lem}

\begin{lem}\label{osserv}
$\mathbb V \left(T_{X^{1:n}},P_X \right) \leq \mathbb V(T_{X^{1:n} Y^{1:n}},P_{XY})$.
\end{lem}

The proof of Lemma \ref{osserv} is straightforward and thus omitted.

\subsection{Achievability proof}\label{subsec: proof}
We want to show that the polar coding scheme proposed in Section \ref{sec:polar} achieves empirical coordination. 
Given $\epsilon >0$, we want to prove that: 
$$\lim_{\substack{n \to \infty \\ k \to \infty}} \mathbb P \left\{ \mathbb V\left(T_{S^{1:n}_{1:k+1} X^{1:n}_{1:k+1} Y^{1:n}_{1:k+1} \widehat{S}^{1:n}_{1:k+1}}, P_{SXY\widehat{S}}\right) > \epsilon \right\} =0.$$

\noindent In order to simplify the notation, we set the joint types as
\begin{align*}
T &:=  T_{S^{1:n}_{1:k+1} \widetilde U^{1:n}_{1:k+1} X^{1:n}_{1:k+1} Y^{1:n}_{1:k+1} \widehat{S}^{1:n}_{1:k+1}},\\
T_i &:=  T_{S^{1:n}_{i} \widetilde U^{1:n}_{i} X^{1:n}_{i} Y^{1:n}_{i} \widehat{S}^{1:n}_{i}} \quad i \in [1,k+1].
\end{align*}

\noindent Lemma \ref{coo1} states that for $i \in [1,k]$ and for all $\epsilon_0 >0$, 
$$\lim_{n \rightarrow \infty} \mathbb P \left\{ \mathbb V \left( T_{S^{1:n}_{i} \widetilde U^{1:n}_{i}},P_{SU}\right) >\epsilon_0 \right\} =0.$$
Then, because of Lemma \ref{lem3}, we have that for all $\epsilon' > \epsilon_0$
$$\lim_{n \rightarrow \infty} \mathbb P \left\{ \mathbb V \left( T_{S^{1:n}_{i} \widetilde U^{1:n}_{i} X^{1:n}_{i} Y^{1:n}_{i}},P_{SUXY}\right) >\epsilon'  \right\} =0.$$
We can apply Lemma \ref{lem3} again and add $\widehat{S}$, but since $\widehat{S}$ is generated by $\widehat{U}$ and not by $\widetilde{U}$, we need the conditional probability: $\forall \epsilon > \epsilon' $ for $i \in [1,k]$ we have 
$$\lim_{n \rightarrow \infty} \mathbb P \left\{ \mathbb V \left( T_i,P_{SUXY\widehat{S}}\right) > \epsilon  \Big| \widehat{U}_i^{1:n} = \widetilde{U}_i^{1:n} \right\} =0$$
We can write:
\begin{align*}
&\mathbb P \left\{ \mathbb V \left( T_i,P_{SUXY\widehat{S}}\right) > \epsilon \right\}\\
&=\mathbb P \left\{ \mathbb V \left( T_i,P_{SUXY\widehat{S}}\right) > \epsilon \Big| \widehat{U}_i^{1:n} = \widetilde{U}_i^{1:n} \right\} \mathbb P \left\{\widehat{U}_i^{1:n} = \widetilde{U}_i^{1:n}  \right\}\\
&+ \mathbb P \left\{ \mathbb V \left( T_i,P_{SUXY\widehat{S}}\right) > \epsilon \Big| \widehat{U}_i^{1:n} \neq \widetilde{U}_i^{1:n} \right\} \mathbb P \left\{\widehat{U}^{1:n}_i \neq \widetilde{U}^{1:n}_i  \right\}.
\end{align*}
Note that the last term tends to 0 since $\widetilde{U}^{1:n}$ is equal to $\widehat{U}^{1:n}$ with high probability because of \eqref{eq: ar}. Hence for $i \in [1,k]$ we have
$$\lim_{n \rightarrow \infty} \mathbb P \left\{ \mathbb V \left( T_i,P_{SUXY\widehat{S}}\right) > \epsilon \right\} =0.$$
The convergence in probability of $T$ to  $P_{S U XY\widehat{S}}$ follows from the convergence in probability of $T_i$ to $P_{S U XY\widehat{S}}$ for $i \in [1,k]$ (coordination in the first $k$ blocks). In fact, observe that by Lemma \ref{vsp},
$$\mathbb V  \left(  T,  P_{S U XY\widehat{S}}\right)
 \leq \frac{1}{k+1} \sum_{i=1}^{k+1} \mathbb V\left(T_i,P_{SU XY\widehat{S}}\right).$$
\noindent This implies that:
\begin{align}\label{eq: e}
 \begin{split}
& \mathbb E_{T} \left[ \mathbb V  \left(  T,  P_{SU XY\widehat{S}}\right) \right] \\
& \leq \frac{1}{k+1} \sum_{i=1}^{k+1}\mathbb E _{T}\left[ \mathbb V\left(T_i,P_{S U XY\widehat{S}}\right) \right].
 \end{split}
\end{align}
The right hand side in \eqref{eq: e} goes to zero since: 
\begin{itemize}
\item for $i \in [1,k]$ we already have the convergence in probability of $\mathbb V\left(T_i,P_{SU XY\widehat{S}}\right)$ to zero, therefore the convergence in mean since
$\mathbb V\left(T_i,P_{SU XY\widehat{S}}\right)$ is bounded for all $i$;
\item for $i=k+1$, since $T_{k+1}$ and $P_{SUXY\widehat{S}}$ are probability distributions, $\mathbb V \left( T_{k+1},P_{SUXY\widehat{S}}\right) \leq 2.$ For $k$ large enough $2/(k+1)$ goes to zero, then $\mathbb E [2]/(k+1)=2/(k+1)$ goes to zero and empirical coordination still holds.
\end{itemize}
Then, the left hand side in \eqref{eq: e} goes to zero and because convergence in mean implies convergence in probability, we have the convergence in probability of $\mathbb V  \left(  T,  P_{S U XY\widehat{S}}\right)$ to zero. To complete the proof we recall that because of Lemma \ref{osserv}, $ \mathbb V\left(T, P_{SU XY\widehat{S}}\right) < \epsilon$ implies that
$$\mathbb V\left(T_{S^{1:n}_{1:k+1} X^{1:n}_{1:k+1} Y^{1:n}_{1:k+1} \widehat{S}^{1:n}_{1:k+1}}, P_{SXY\widehat{S}}\right) < \epsilon.$$

\appendix \label{sec: varie}

\subsection{Proof of Lemma \ref{coo1}}
For all $\epsilon_0 >0$, we define
  \begin{align*}
  &\mathcal{T}_{\epsilon_0}\left( P_{SU} \right) := \left\{ (S^{1:n},U^{1:n}) \big| \mathbb{V}\left( T_{S^{1:n} U^{1:n}}, P_{SU} \right) \leq \epsilon_0\right\}\\
  &\mathbb P_{P_{SU}} \left\{  (s^{1:n},u^{1:n}) \in \mathcal{T}_{\epsilon_0}\left( P_{SU} \right) \right\}:=\\
  & \quad \sum_{s^{1:n},u^{1:n}} P_{S^{1:n}U^{1:n}} \left(s^{1:n},u^{1:n} \right)\mathds{1}\left\{ (s^{1:n},u^{1:n}) \in  \mathcal{T}_{\epsilon_0}\left( P_{SU} \right) \right\}.
  \end{align*}
Note that $\lim_{n \rightarrow \infty} \mathbb P_{P_{SU}} \left\{  (s^{1:n},u^{1:n}) \in \mathcal{T}_{\epsilon_0}\left( P_{SU} \right) \right\} =1$.

Let $i \in [1,k]$, we have:
\begin{align*}
& \mathbb P_{P_{S \widetilde U}} \left\{ \mathbb{V}\left( T_{S^{1:n}_i U^{1:n}_i}, P_{SU}\right) > \epsilon_0\right\}\\
& = \sum_{s^{1:n},u^{1:n}} P_{S^{1:n}_i \widetilde U^{1:n}_i} \left(s^{1:n},u^{1:n} \right) \mathds{1}\left\{ (s^{1:n},u^{1:n}) \notin  \mathcal{T}_{\epsilon_0}\left( P_{SU} \right) \right\}\\
  & =  \sum_{s^{1:n},u^{1:n}} ( P_{S^{1:n}_i  \widetilde U^{1:n}_i} \left(s^{1:n},u^{1:n} \right)-  P _{S^{1:n}U^{1:n}} \left(s^{1:n},u^{1:n} \right) \\
  & +  P _{S^{1:n}U^{1:n}} \left(s^{1:n},u^{1:n}\right) ) \mathds{1}{\left\{ (s^{1:n},u^{1:n}) \notin  \mathcal{T}_{\epsilon_0}\left( P_{SU} \right) \right\}}\\
  & \leq \mathbb{V} ( P_{S^{1:n} \widetilde U^{1:n}}, P_{S^{1:n}U^{1:n}}) + \mathbb P_{P_{SU}} \left\{  (s^{1:n},u^{1:n}) \notin \mathcal{T}_{\epsilon_0}\left( P_{SU} \right) \right\}
\end{align*}
which tends to 0 thanks to a typicality argument and the following result.

\begin{lem}\label{dis}
For any $i \in [1,k]$, let $\delta_n = 2^{-n^ {\beta}}$ for some $0<\beta<1/2$
 $$\mathbb{V} \left( P_{U^{1:n}S^{1:n}}, {P}_{\widetilde U_i^{1:n}S_i^{1:n}} \right) \leq \sqrt{2 \log2} \sqrt{n \delta_n}.$$
 \begin{IEEEproof}
 We have
  \begin{align*}
& \mathbb{D}\left( P_{U^{1:n}S^{1:n}} \Big\Arrowvert P_{\widetilde U_i^{1:n}S_i^{1:n}} \right)  {\overset{{(a)}}{=}}
\mathbb{D}\left( P_{V^{1:n}S^{1:n}} \Big\Arrowvert {P}_{\widetilde V_i^{1:n}S_i^{1:n}}  \right)\\
   &  {\overset{{(b)}}{=}} \mathbb{D}\left( P_{V^{1:n}|S^{1:n}} \Big\Arrowvert {P}_{\widetilde V_i^{1:n}|S_i^{1:n}} \Big| P_{S^{1:n}}\right)\\
   &  {\overset{{(c)}}{=}} \sum_{j =1}^n \mathbb D \left( P_{V^j|V^{1:j-1} S^{1:n} }\Big\Arrowvert {P}_{\widetilde V_i^j| \widetilde V_i^{1:j-1} S_i^{1:n}} \Big| P_{V^{1:j-1} S^{1:n}}  \right)\\
   &  {\overset{{(d)}}{=}}\sum_{j \in A_1 \cup A_2} \mathbb D \left(  P_{V^j|V^{1:j-1} S^{1:n} }\Big\Arrowvert {P}_{\widetilde V_i^j| \widetilde V_i^{1:j-1} S_i^{1:n} } \Big| P_{V^{1:j-1} S^{1:n}}\right)\\
   &  {\overset{{(e)}}{=}} \sum_{j \in A_1 \cup A_2} \left( 1- H(V^j \mid V^{1:j-1}S^{1:n}) \right) {\overset{{(f)}}{<}} \delta_n \lvert \mathcal V _{V\mid S}\rvert 
   \leq n \delta_n,
  \end{align*}
where $(a)$ comes from the invertibility of $G_n$, $(b)$ and $(c)$ come from the chain rule, $(d)$ comes from \eqref{eq: p1} and \eqref{eq: p2}, 
$(e)$ comes from the fact that the conditional distribution $P_{\widetilde V_i^j|\widetilde V_i^{1:j-1}S_i^{1:n}}$ is uniform for $j$ in $A_1$ and $A_2$ and $(f)$ from \eqref{eq: hv}. Then, the proof is completed using Pinsker's inequality. \end{IEEEproof}
\end{lem}
\subsection{Proof of Lemma \ref{lem3}}
We have:
\begin{footnotesize}
\begin{align*}
& \mathbb P \left\{\mathbb V\left(T_{A^{1:n} B^{1:n}},P_{AB}\right)> \epsilon'\right\} \leq \mathbb P \left\{ \mathbb V \left(T_{A^{1:n}}, P_A \right) > \epsilon \right\} +\\
& \mathbb P \left\{\mathbb V \left(T_{A^{1:n}}, P_A\right)\leq \epsilon \right\}  \mathbb P \left\{\mathbb V \left(T_{A^{1:n} B^{1:n}},P_{AB}\right)> \epsilon' \big| \mathbb V\left(T_{A^{1:n}}, P_A\right)\leq \epsilon\right\}.
\end{align*}
\end{footnotesize}
Then as $n$ goes to infinity, the first term tends to zero by the conditional typicality lemma \cite{csiszar2011information} and the second tends to zero by hypothesis.

\subsection{Proof of Lemma \ref{vsp}}
The statement follows from the inequalities:
\begin{footnotesize}
\begin{align*}
& \big\lvert T_{\left(X^n,\widetilde X^n\right)} (x) - P(x) \big\rvert \\
& = \Bigg\lvert  \frac{1}{2} \sum_{i=1}^n \left( \frac{ \mathds 1 \{ X^i=x\}}{n} + \frac{ \mathds 1 \{\widetilde X^i=x\}}{n}\right) - \frac{P(x)}{2} - \frac{P(x)}{2} \Bigg\rvert\\
& \leq    \frac{1}{2} \Bigg\lvert \sum_{i=1}^n  \frac{\mathds 1 \{X^{i}=x\} }{n} -P(x)\Bigg\rvert  +  \frac{1}{2} \Bigg\lvert \sum_{i=1}^n   \frac{ \mathds 1 \{\widetilde X^i=x\}}{n} -P(x) \Bigg\rvert.
\end{align*}
\end{footnotesize}

\begin{footnotesize}
\bibliographystyle{IEEEtran}
\bibliography{mybib}
\end{footnotesize}
\end{document}